\documentclass[pdflatex,sn-mathphys-num]{sn-jnl}

\usepackage{graphicx}%
\usepackage{multirow}%
\usepackage{amsmath,amssymb,amsfonts}%
\usepackage{amsthm}%
\usepackage{mathrsfs}%
\usepackage[title]{appendix}%
\usepackage{xcolor}%
\usepackage{textcomp}%
\usepackage{manyfoot}%
\usepackage{booktabs}%
\usepackage{algorithm}%
\usepackage{algorithmicx}%
\usepackage{algpseudocode}%
\usepackage{listings}%

\usepackage[T1]{fontenc}
\usepackage{lmodern}

\theoremstyle{thmstyleone}%

\theoremstyle{thmstyletwo}%

\theoremstyle{thmstylethree}%

\newcommand\MOD[1]{{\color{black}#1}}
\newcommand\rev[1]{{\color{black}#1}}

\begin{document}

\title{\centering The dimensions of accessibility:\\proximity, opportunities, values}

\author*[1,2]{\fnm{Matteo} \sur{Bruno}}\email{matteo.bruno@sony.com}
\author[1,2]{\fnm{Bruno} \sur{Campanelli}}
\author[3,4,1,2]{\fnm{Hygor P.} \sur{M. Melo}}
\author[5,1,2]{\fnm{Lavinia} \sur{Rossi Mori}}
\author[1,5,2,6]{\fnm{Vittorio} \sur{Loreto}}

\small{
\affil[1]{\orgname{Sony Computer Science Laboratories - Rome}, Joint Initiative CREF-SONY, Centro Ricerche Enrico Fermi, \orgaddress{\street{Via Panisperna 89/A}, \postcode{00184}, 
\city{Rome}, \country{Italy}}}
\affil[2]{\orgname{Centro Ricerche Enrico Fermi (CREF)}, \orgaddress{\street{Via Panisperna 89/A}, \postcode{00184}, \city{Rome}, \country{Italy}}}
\affil[3]{\orgname{Postgraduate Program in Applied Informatics, 
Univ. of Fortaleza}, \orgaddress{\street{60811-905}, \city{Fortaleza}, \state{CE}, \country{Brazil}}}
\affil[4]{\orgname{Núcleo de Ciência de Dados e Inteligência Artificial (NCDIA), Univ. of Fortaleza}, \orgaddress{\postcode{60811-905}, \city{Fortaleza}, \state{CE}, \country{Brazil}}}
\affil[5]{\orgname{Sapienza Univ. of Rome}, \orgdiv{Physics Dept}, \orgaddress{\street{Piazzale A. Moro, 5}, \postcode{00185}, \city{Rome}, \country{Italy}}}
\affil[6]{\orgname{Complexity Science Hub}, \orgaddress{Metternichgasse 8}, \postcode{A 1030} \city{Wien}, \country{Austria}}
}



\abstract{Accessibility is essential for designing inclusive urban systems. However, the attempt to capture the complexity of accessibility in a single universal metric has often limited its effective use in design, measurement, and governance across various fields. Building on \rev{previous work by Bertolini and} by Levinson and Wu, we emphasise that accessibility consists of three key dimensions. Specifically, we introduce a conceptual framework that defines accessibility through three main dimensions: \textit{Proximity} (which pertains to active, short-range accessibility to local services and amenities), \textit{Opportunity} (which refers to quick access to relevant non-local resources, such as jobs or major cultural venues), and \textit{Value} (which encompasses the overall quality and personal significance assigned to specific points of interest). While it is generally beneficial to improve accessibility, different users and contexts present unique trade-offs that make a one-size-fits-all solution neither practical nor desirable. Our framework establishes a foundation for a quantitative and integrative approach to modelling accessibility. It considers the complex interactions among its various dimensions and facilitates more systematic analysis, comparison, and decision-making across diverse contexts.}


\maketitle


\section{Introduction}\label{sec1}

Accessibility is a broad term that encompasses various meanings, relating to the ability to reach preferred destinations. It is an interdisciplinary concept that plays a central role in fields such as geography~\cite{giuliano2017geography}, urban planning~\cite{levine2019mobility}, social science~\cite{cass2005social}, economics~\cite{fujita2001spatial}, and engineering~\cite{ortuzar2002modelling}. In these areas, accessibility can refer to physical mobility, social or digital inclusion, availability of services, economic opportunities, and participation in society. Each field offers its own definitions and priorities, leading to different frameworks and interpretations. Despite this diversity, the core idea remains consistent: accessibility involves the relationship between individuals and the activities, places, or services they need to access within a system. The complexity of this concept makes accessibility both significant and challenging to define, highlighting the need for a structured and integrated framework.



\rev{The complex concept of accessibility has been explored by researchers} across various fields over time. One of the earliest attempts to formalise this idea was made by Hansen~\cite{hansen1959accessibility}, who defined accessibility as a function based on the number of opportunities available in a specific location divided by the time required to reach them. Hansen proposed a sort of gravity law for accessibility, suggesting that areas with more opportunities tend to attract more people, with this attraction decreasing as the distance from those locations increases. \rev{Analysing Transit-oriented development (TOD), the concept of node-place access was developed by Bertolini~\cite{bertolini1996nodes,bertolini1999spatial,wu2023node}, introducing a notion of far and proximal access. In this seminal work, places close to rail stations and with therefore good transit accessibility are called nodes, whereas places with high proximity and high density are called places, separating accessibility to services in two separate dimensions.} In more recent years, Levinson and Wu have sought to unify different approaches to quantifying accessibility to develop a more comprehensive theory and measurement framework~\cite{levinson2020towards}. Their work reviews previous research on accessibility from various perspectives, summarising what can be measured and the methodologies used. This significant effort aims to consolidate multiple approaches into a single guide for measuring accessibility by examining the essential questions of "where," "what," "when," "why," and "how" to quantify access. They propose a generalised measure of accessibility. Additionally, other studies have reviewed the literature on accessibility metrics, highlighting both standard measurements and innovative approaches to understanding accessibility~\cite{handy1997measuring,geurs2004accessibility,elgeneidy2006access,wu2020unifying}.

Accessibility has inspired various urban models, often focused on optimal standards for different transportation modes. These models include walkable cities~\cite{speck2012walkable}, transit-oriented development, car-centric urban layouts, and more recently, concepts like chrono-urbanism and the 15-minute city~\cite{moreno2021introducing}. Each model reflects specific assumptions about mobility, infrastructure, spatial distribution, and equity. However, what is considered "optimal" accessibility in one context may not apply in another, making it challenging to achieve generalised accessibility optimally.

\rev{In this work, we move beyond reviewing or refining existing metrics and instead}
challenge the concept of accessibility itself by distinguishing between different types of access needs\rev{, offering a quantitative framework that builds on the node-place theory~\cite{bertolini1996nodes} while adding a new dimension to it}. We argue that the nature of access varies depending on the type of amenity one needs to reach and demonstrate how different metrics can capture these distinct types of access. We differentiate between access to everyday services and access to key opportunities, proposing a novel framework for understanding accessibility. This framework quantifies the various dimensions of access using distinct metrics. We also apply these metrics and framework to real-world cases, analysing accessibility patterns in a set of cities from different countries. Our findings show that this framework can capture complex spatial patterns, highlighting specific areas' infrastructure and service needs. This information can be used to improve accessibility in isolated urban zones. Finally, we discuss the difficulties in achieving perfect optimisation of accessibility, given the complexity of constraints and the inherent geographic inequalities in different regions.

\section{The dimensions of accessibility}

We start from the definition of accessibility, often referred to as \textit{the ease of reaching valued destinations}, for instance in Levinson and Wu~\cite{levinson2020towards}. This simple definition hides a lot of subtleties: how to define the \textit{ease}? What does \textit{reaching} mean? How to measure the \textit{value}? Which \textit{destinations} can be considered?

The current work does not aim at clarifying these questions, which are thoroughly examined in~\cite{levinson2020towards} and elsewhere. We would rather focus on understanding which elements, or \textit{building blocks}, make a completely satisfying description of accessibility. A \textit{complete neighbourhood}, theoretically, should allow the residents to access everything they need. However, this might not be possible depending on the complexity of the amenities that one needs to reach. The case of access to jobs here is paramount. There cannot reasonably be enough density to have a research laboratory within any 15-minute walk, an international law firm, or a university. There are other examples: a specific martial art gym, a monument, a Michelin-starred restaurant, and a medieval museum. Conversely, it is also complex to imagine a housing market that allows people to move freely and quickly close to their job location, and even if such a market existed, households can host people with jobs in different areas. People can get used to a local supermarket even if it's not their favourite brand, or have coffee at their neighbourhood café and not at the fancy downtown one. However, renouncing a job just because it is not in the local area is another question. 
\begin{figure}[h]
\centering
\includegraphics[width=0.4\textwidth]{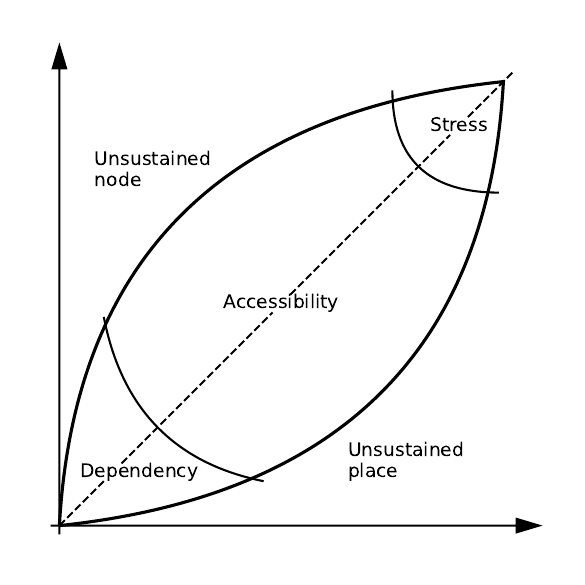}
\includegraphics[width=0.49\textwidth]{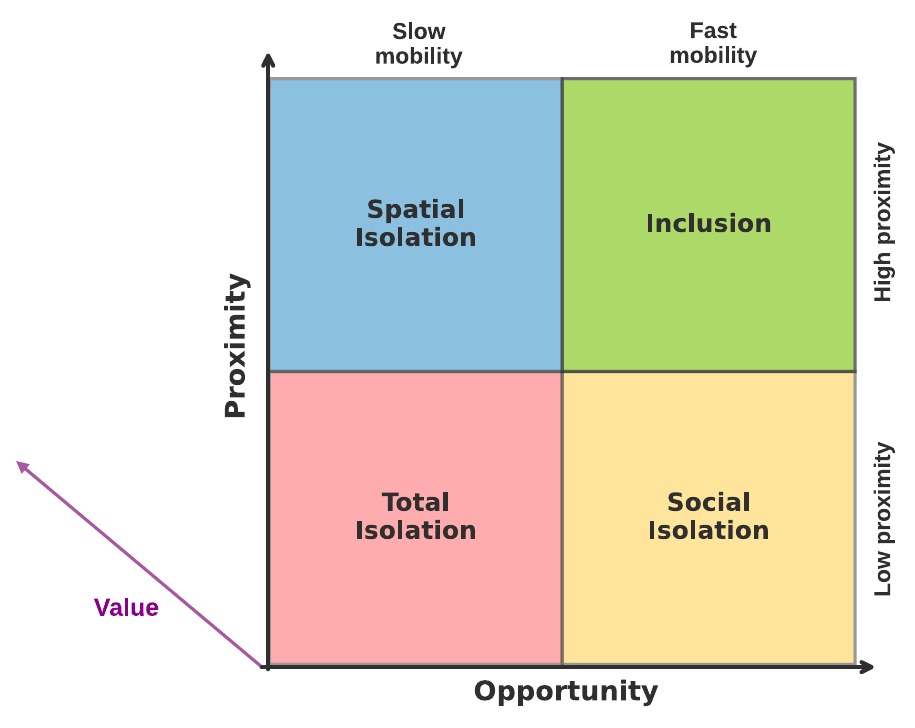}
\caption{\textbf{Dimensions of accessibility.} \rev{Left: The original node-place diagram in Bertolini, 1999~\cite{bertolini1999spatial}}. \MOD{Right: Graphic illustration of} the three dimensions of accessibility discussed in this paper: Proximity, Opportunity, and, still in a draft way, Value. These three dimensions are not necessarily correlated. For example, areas with high proximity access but low access to opportunities can be spatially isolated from the rest of the city (Spatial Isolation, located at the top left). Conversely, regions with low proximity access but high access to opportunities may be connected to the city via transport infrastructure but lack a vibrant social neighbourhood life (Social Isolation, found at the bottom right). Inclusion and Total Isolation are terms that are clear in their meaning. The Value dimension further complicates the situation because it can disrupt the above-mentioned categories. For instance, good access may hold little value if the accessible places are not considered valuable.
}\label{fig1:building_blocks}
\end{figure}
Therefore, we propose two kinds of accessibility: a local one (Proximity) and a more global one (Opportunity). \rev{The two dimensions aim at capturing different accessibility metrics for unifying the concept of node and place in the node-place theory~\cite{bertolini1999spatial}.} In the upcoming discussion, we will refer to Fig.~\ref{fig1:building_blocks} that describes accessibility in 2+1 dimensions: Proximity, Opportunity and, in a less precise way, Value.

\subsection{Proximity access and Opportunity access}

Not all services have the same functions and frequency of use, and therefore, accessing different amenities can be described on various scales of distance (or time). Services that people need for everyday needs, such as \textit{grocery stores}, \textit{pharmacies}, \textit{banks}, or \textit{restaurants}, can be studied through the lens of \textbf{proximity}. \MOD{The level of local proximity experienced in a neighbourhood can be reproduced regardless of a city’s overall size}, starting from the centre of a small town and moving up to a vibrant neighbourhood in a megalopolis. Why then do people move from rural areas or small towns to big cities? It is the need for \textbf{opportunity} fuels people's movement towards big cities. Opportunities in cities are on a different spatial scale compared to proximity services: we need fast mobility to be able to reach a particular kind of \textit{job}, a city \textit{stadium}, a large opera \textit{theatre} or \textit{museum} and a historical \textit{monument}~\cite{hill2024beyond}. The two scales of proximity and opportunity are only marginally correlated. For example, \MOD{one can only reach a stadium in terms of proximity if they live near one. Also, one would consider themselves very lucky to have found a suitable job within a 15-minute walk from their home in a city like Tokyo, which still has a very good proximity accessibility~\cite{brunoUniversalFrameworkInclusive}}. Therefore, we argue that these two accessibility scales could ideally be \textbf{independent dimensions} of accessibility. We could easily think of measuring the two dimensions using independent metrics. For instance, if the {\em proximity} $A_i$ of a location $i$ is calculated as the number of opportunities within 15 minutes, 
\begin{equation}
    A^{prox}_i = \sum_j O_{ij}\cdot \Theta_{0<t<15}\;,
\end{equation}
and the {\em opportunity} is calculated as the number of opportunities between 15 and 60 minutes, 
\begin{equation}
    A^{opp}_i = \sum_j O_{ij}\cdot \Theta_{15<t<60}\;,
\end{equation}
then these two metrics will \textit{theoretically} be independent, meaning that their values can be arbitrarily different if the services are located primarily on proximity or mostly remotely. Here $O_{ij}$ is the number of opportunities in $j$ seizable from $i$ and the function $\Theta_{t_1<t<t_2}$ filters only the opportunities between $t_1$ and $t_2$.

In practice, proximity and opportunity are often correlated, with population density as fertile ground for both. Services, infrastructure, and populations are typically spatially related. Proximity services require densely populated areas to be viable and economically successful, with enough clients in their catchment area to thrive. Similarly, mobility—particularly public transport—demands a specific density to be profitable~\cite{cervero1997travel,guerra2011cost}. For example, metro stations need at least 7,000 people or jobs per square kilometre to remain sustainable with subsidies; even higher densities are required for profitability. \rev{This argument is similar to the node-place independent characterisation: the nodes and places can be theoretically uncorrelated, as a railway station can have low proximity and high mobility whereas a place can have low mobility, although these two kinds of accessibility often attract each other as highly reachable places will attract activities and services and dense areas will push for faster transportation options.}

The two dimensions of accessibility can often confuse the general public. The concept of chrono-urbanism has led many people to conflate terms like “15-minute city”, “car-free areas”, “low traffic zones”, and “30 km/h zones”. Logically speaking, a perfect 15-minute city could still suffer from congested roads, slow public transportation, or inefficient metro systems. This is commonly observed in cities with dense areas that have recently experienced rapid car-oriented growth, such as Rome, Athens, Bogotá, and Fortaleza. Conversely, cities like Oxford and Marseille~\cite{brunoUniversalFrameworkInclusive} may have good proximity services but less effective public transport.

Moreover, suburban areas designed for car use might offer fast traffic but lack proximity opportunities, as seen in lower-density cities in the U.S. (such as San Antonio and Oklahoma City), Oceania (like Melbourne and Auckland), and Northern Europe (Milton Keynes, Stockholm). Even peripheral residential areas with good public transit connections—such as Fukuoka and Chengdu—can suffer from this issue.

Recent studies suggest that actual 15-minute areas can help local populations decrease their need for movement~\cite{abbiasov2022quantified}; however, this mainly pertains to short-range services. Consequently, long-distance commutes and access to remote opportunities remain unaddressed. More recent studies~\cite{marzolla2024compact,marzolla2025proximity} showed that per capita transportation emissions correlate with both the proximity of services and the size of cities, indicating that larger areas increase the need for movement. This further underscores the necessity of distinguishing between accessing close and remote services. 

Therefore, long-distance access must be supported by sustainable and efficient transportation options. Metro lines, bus rapid transit, and expansive bicycle networks are essential to reducing car dependency, simplifying remote connections, and improving distant accessibility. In summary, while proximity is necessary, it must be complemented by fast and sustainable transport to reduce emissions~\cite{cervero1997travel,guerra2011cost} effectively.

\subsection{The value}

Beyond distinguishing between {\em proximity} and {\em opportunity}, another critical aspect of accessibility relates to the value attributed to reachable services or points of interest. Even when two neighbourhoods meet the proximity criteria, the practical accessibility perceived by residents can vary significantly based on the quality, desirability, and diversity of the services available. 

\rev{An extension of the two accessibility axes was already proposed in the node-place framework by previous researchers. For instance, Vale et al.~\cite{vale2018extended} proposed the third dimension of "design", while Cao et al.~\cite{cao2020coordination} propose a "ridership" dimension as a third axis. Our approach, where the third dimension should capture the hidden value of the territory, is generalising these proposals: the design might be an element that adds to the value of a place through beauty and functionality; the ridership would be related to how much effectively a node/place is experienced, resulting from the value of its territory and connections.}

The value dimension includes factors such as the quality of service provision, reputation, variety, and how well local offerings align with residents' specific needs, preferences, and expectations. For example, school access is not just about being close by; it also involves educational quality, extracurricular activities, and perceived social value. Similarly, two neighbourhoods might have many restaurants within walking distance, but they can differ significantly regarding culinary variety, pricing, and overall dining quality.

The spatial distribution of service quality is usually uneven and is systematically influenced by socio-economic factors. Higher-quality establishments often cluster in affluent neighbourhoods, creating disparities in adequate accessibility, sometimes called "quality deserts" in lower-income areas~\cite{Morland2002}. Empirical studies illustrate this clearly regarding food accessibility: economically disadvantaged neighbourhoods may have a surplus of low-quality, unhealthy food outlets while lacking access to fresh produce and quality dining options~\cite{Block2004,GarcaBulleBueno2024}.

Recognising value as a crucial aspect of accessibility underscores the importance of equitable urban planning. It must go beyond merely ensuring the physical presence of services. Policymaking and urban governance should actively pursue strategies to distribute high-quality services more equitably across neighbourhoods, providing physical proximity or opportunity and genuine access to valued, high-quality amenities.

\subsection{Isolation}

In our accessibility framework, we separate the dimensions of accessibility in proximity and opportunity. Logically, the effects of the presence and lack of proximity and opportunity must also be divided. The lack of accessibility can be generally called by many names, depending on the perspective. Here we use the word \textit{isolation}, but words like exclusion and even segregation could also fit, when focusing on specific accessibility like jobs or public services.

When a territory does not have either of the two types of accessibility, we can call this \textit{total isolation}. In this case, the place is not very well connected through public transportation to the main activities of the cities and does not have a social fabric in its local neighbourhood. A car is probably essential to live here.

In the case of high proximity and low opportunity, there is a spatial mismatch~\cite{kain2004pioneer}, meaning that a person living in such a place might have a local community and social connections, but will be disconnected from less local opportunities, including jobs (especially high-skilled job opportunities). This notion is what we call \textit{spatial isolation} (see top-left of Fig.~\ref{fig1:building_blocks}).

When, instead, a neighbourhood has good public transport and low social structure, with the absence of a square and places for social aggregation, we can name this \textit{social isolation} (see bottom-right of Fig.~\ref{fig1:building_blocks}). A person living in such a place could have a competitive job in a central place in the city and might be able to travel to main cultural hubs, but will lack proximity services and a sense of belonging to a local community.

\MOD{When choosing thresholds of proximity and opportunity below which we should consider an area as isolated, there is clearly a degree of arbitrariness. In the results section
we use city-specific thresholds corresponding to the median values of proximity and opportunity experienced by the city's population.}

The isolation types just described are not automatically undesirable for all. Some people still prefer to live in suburbs, affording a larger house and renouncing local and global opportunities, for instance, growing a family in a larger space or enjoying quieter surroundings. Age, culture, needs and personal preferences make each person's desirability scale very specific. Too many opportunities could also be detrimental for some~\cite{schwartz2015paradox}, and a very high density could lower the care for the local environment.

After all these considerations about accessibility, we still should put value in the equations. Even if one has the best proximity possible to local services, if the available restaurants, schools, and banks offer bad service or mediocre quality, then this type of proximity would still be rather disappointing, especially compared to the expected standards of the rest of the population. This is the case, for instance, of tourist areas, where prices are inflated and quality can be disappointing by the residents' average standards.

\subsection{Optimising accessibility globally and locally}

While it is theoretically possible to increase accessibility by adding services and infrastructure, achieving equitable configurations across population distributions, infrastructure, and services is much more challenging. A city has both a geometric and functional structure, with a defined centre and periphery. Depending on the landscape and the distribution of infrastructure and services, a city's functional centre may not align with its geometric center~\cite{sun2016identifying,fanelli2025revealing}. Consequently, people tend to live closer to the centre, while others reside in more peripheral areas.

Moreover, the opportunities available in a city are not absolute. What might be considered the best opportunity in a small town may only qualify as a local option in a mid-sized city. In contrast, the most desirable area in a city with one million residents may pale compared to the opportunities in a major megalopolis. Therefore, optimising access to opportunities is a complex challenge, affecting society and individuals~\cite{gastner2006optimal,xu2020deconstructing,brunoUniversalFrameworkInclusive}. 

Various redistribution algorithms have been proposed to reduce inequalities. While these approaches may help decrease the gap between a vibrant urban core and underserved peripheral areas by improving access to basic services, rare opportunities will still tend to be concentrated among a few privileged individuals. Additionally, attractive amenities often draw more amenities due to the high concentration of people they attract, creating a "rich-get-richer" dynamic. Although policies can help mitigate this effect, some spatial and social diversity will inevitably remain.

As the third dimension of our framework, value is statistically challenging to equalise. Typically, the perceived quality of restaurants varies significantly. The best restaurants in a city are relatively few, and if they thrive in the market, they can choose prime locations to attract the most customers. This often means establishing themselves in the city centre or in areas that are easily accessible, even from long distances. Consequently, the locations with the highest concentration of restaurants also tend to attract the highest quality establishments. Even if high-quality restaurants were randomly distributed, the most accessible areas would still end up housing most of them simply because of the larger number of restaurants in those locations.

\section{Methods}\label{sec11}

\subsection{Data}

The cities in this study are defined through their administrative boundaries, each obtained from the respective municipality's open data portal. Public transport schedules were collected in GTFS format from the websites of the relevant transport agencies in May 2025. To the best of our knowledge, the selected cities represent all cases for which up-to-date GTFS datasets were available for all agencies operating within the municipal territory. Points of Interest (POIs), used here as a proxy for opportunities, were extracted from OpenStreetMap~\cite{OpenStreetMap} data. Population data were downloaded from WorldPop~\cite{worldpop}. A 100-meter resolution population density grid was used, adjusted to match municipal population estimates provided by the United Nations~\cite{bondarenko2020census}.

\subsection{Accessibility metrics}
Each city is covered with a hexagonal grid using the H3 geospatial indexing system~\cite{h3} with resolution 9, corresponding to hexagons with sides of approximately \MOD{200 m}; all H3 hexagons whose centroids fall within the city boundaries are considered part of the city. The travel
times by foot between all centroid pairs are computed using the Open Source Routing Machine (OSRM)~\cite{luxen2011osrm}, with OpenStreetMap data as the underlying road network source. The walking times are then combined with the public transport schedules via the Connection Scan Algorithm to obtain travel times via public transport, under the constraint that travellers can walk for a maximum of 15 minutes at a speed of \MOD{5 km/h} to move between different public transport stops, from the origin centroid to a stop, or from a stop to the destination centroid. As a simplifying assumption, all POIs contained in a destination hexagon $d$ are considered accessible after travelling for a time $t$ if and only if $d$'s centroid is reachable.


\textit{Proximity.} We define a hexagon's $\lambda$ proximity value as
\begin{equation}
    A^\text{prox}_\lambda = \int_0^\infty dt\,\rho_\lambda^{\text{foot}}(t)f^\text{foot}(2t)\;,
    \label{eq:proximity}
\end{equation}
where $\rho_\lambda^{\text{foot}}(t)$ is the number of POIs that can be reached by starting from $\lambda$ and walking for a time $t$, and $f^\text{foot}(t)$ is a utility function representing the diminishing value of POIs that are harder to reach:
\begin{equation}
    f^\text{foot}(t) = \frac{1}{\tau}e^{-t/\tau} \;,\;\;\tau = 60\,\text{min}.
\end{equation}
Note that in Eq. (\ref{eq:proximity}) the utility function is calculated in $2t$ to account for the return trip.


\textit{Opportunity.} While proximity measures the ability of a person residing in a hexagon to find services and amenities within walking distance, opportunity is concerned with the ease of "exploring the city", even relatively far from one's home: 
\begin{equation}
    A^\text{opp}_\lambda = \left\langle\int_0^\infty dt\, \rho_\lambda^\text{pub}(t_0, t)f^{\text{pub}}(2t)\right\rangle_{t_0}\;.
\end{equation}
As the access to services and amenities not in close proximity is typically accomplished by using public transport, we now use the number of POIs reachable by public transport $\rho_\lambda^\text{pub}(t_0, t)$. This element introduces a dependency on the journey's starting time $t_0$, as public transport schedules vary throughout the year and the time of day. To eliminate this, for each GTFS feed, we select the day with the highest count of public transport trips, and average the calculation over several starting times (the beginning of every hour between 6 AM and 10 PM). The utility function is now given by the probability of having a journey of duration $t$ as derived in~\cite{kolblEnergyLawsHuman2003}:
\begin{equation}
    f^\text{pub} = N\exp{\left(-\frac{\alpha T}{t}\right)} \exp{\left(-\frac{\beta t}{T}\right)}\;\;,\;\;\alpha=0.2,\;\beta=0.7,\;T=67\text{min}\;,
\end{equation}
where $N$ is a normalisation constant such that the integral between 0 and $\infty$ is 1. A similar approach has been employed in~\cite{biazzo2019general}, with reachable area and people in place of opportunities. This function decays spatially in a much slower fashion than $f^\text{foot}(t)$ due to the higher average speed of public transport compared to walking, and vanishes exponentially for small $t$. \MOD{As shown in~\cite{biazzo2019general}, the qualitative behaviour of the resulting accessibility indicators is robust to the specific choice of decay kernel, provided that the same functional form is used consistently for both $f^\text{foot}(t)$ and $f^\text{pub}(t)$. However, alternative kernels can be adopted to better match specific empirical or policy contexts.}

\MOD{The formulation above can be readily extended to recently proposed 3D computations of accessibility times~\cite{ng20253d,CHEN2026106516} by employing a 3D network in the calculation of $\rho_\lambda(t)$, although in this work we rely on the conventional 2D transport network. Such approaches could refine the estimates by accounting for elevation changes, slopes, or multi-level buildings in dense urban areas.}

\section{Results}\label{sec:results}

We calculate proximity and opportunity metrics for eighteen cities in the world. 
\begin{figure}[h]
\centering
\includegraphics[width=0.99\textwidth]{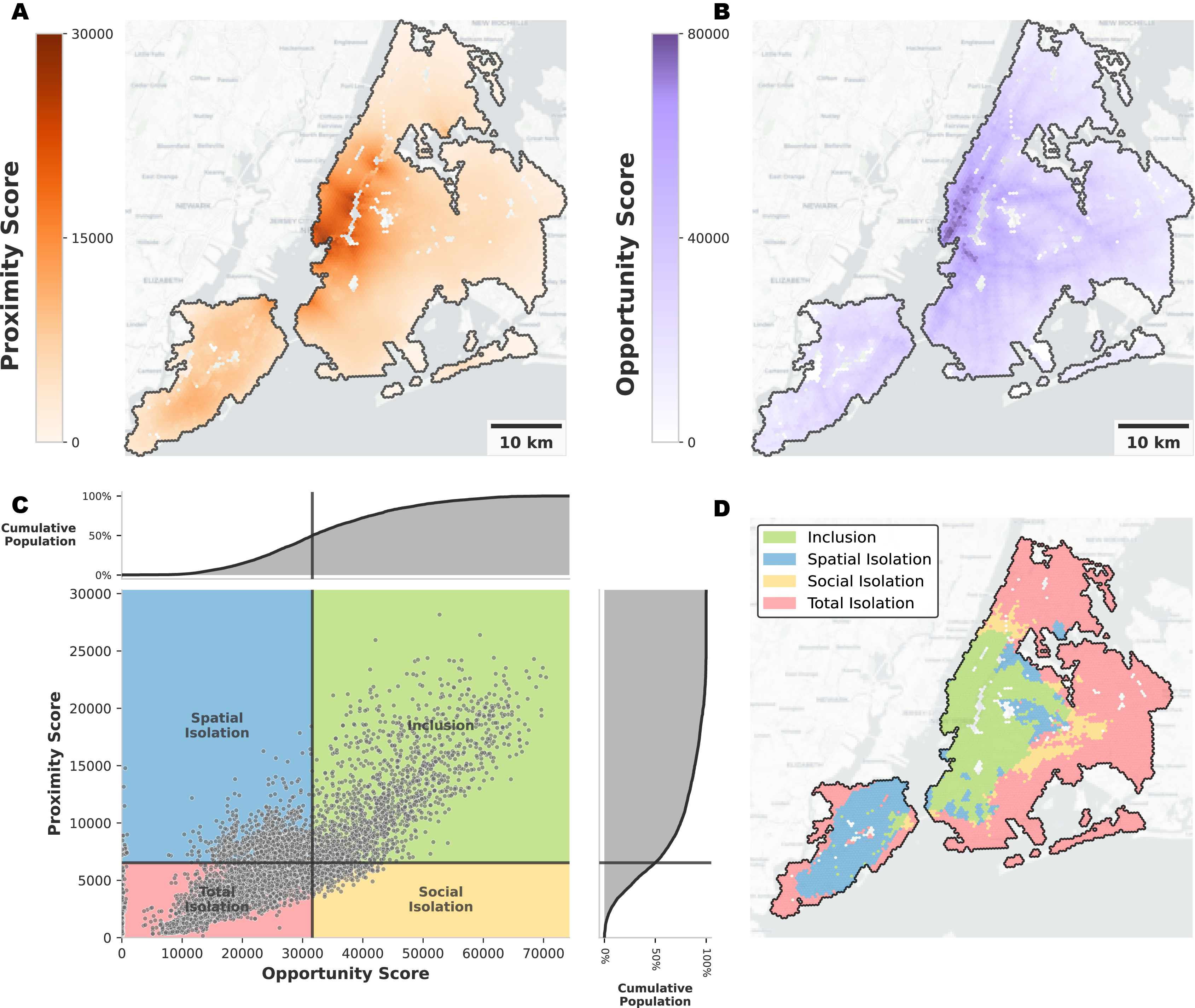}
\caption{\textbf{Diagram with dimensions of accessibility for New York.} (A) and (B) show, respectively, a map of the proximity and opportunity scores for the New York area. (C) represents a scatter plot of the hexagon's values, showing the correlation between the two metrics and the fluctuations and a proposed categorisation of the areas based on their \MOD{median} values.
(D) The Manhattan Island and the Brooklyn area are very well connected in both accessibility dimensions, while the Harlem/South Bronx area lacks proximity. Staten Island, on the other hand, lacks fast opportunity access compared to the rest of the city. 
}\label{fig2:ny_case_study}
\end{figure}
Figure \ref{fig2:ny_case_study} summarises the results for New York. From panels A, B and C, it is evident that there is a positive correlation between proximity and opportunity, \rev{as also found in previous studies~\cite{bertolini1999spatial,wu2023node}}. There is still room for significant fluctuations - Pearson's correlation coefficient between the logarithms of proximity and opportunity is $0.61$, while it ranges between $0.47$ (Zurich) and $0.83$ (Paris) for the other cities. The isolated cluster of points in the bottom-left corner of panel C consists of hexagons from which no public transport stops can be accessed within 15 minutes. Panels C and D are divided into coloured zones according to the \MOD{median} accessibility values. People living in the green zone enjoy above-median accessibility in both dimensions, while people living in the other zones experience below-median accessibility in at least one dimension. \MOD{Across the eighteen cities in our sample, we generally observe that low–proximity, low-opportunity zones concentrate in peripheral belts, although the sharpness of this separation varies from city to city, in line with previous evidence that low-proximity locations are typically found at the urban fringe~\cite{brunoUniversalFrameworkInclusive}.}
\begin{figure}[h]
\centering
\includegraphics[width=0.9\textwidth]{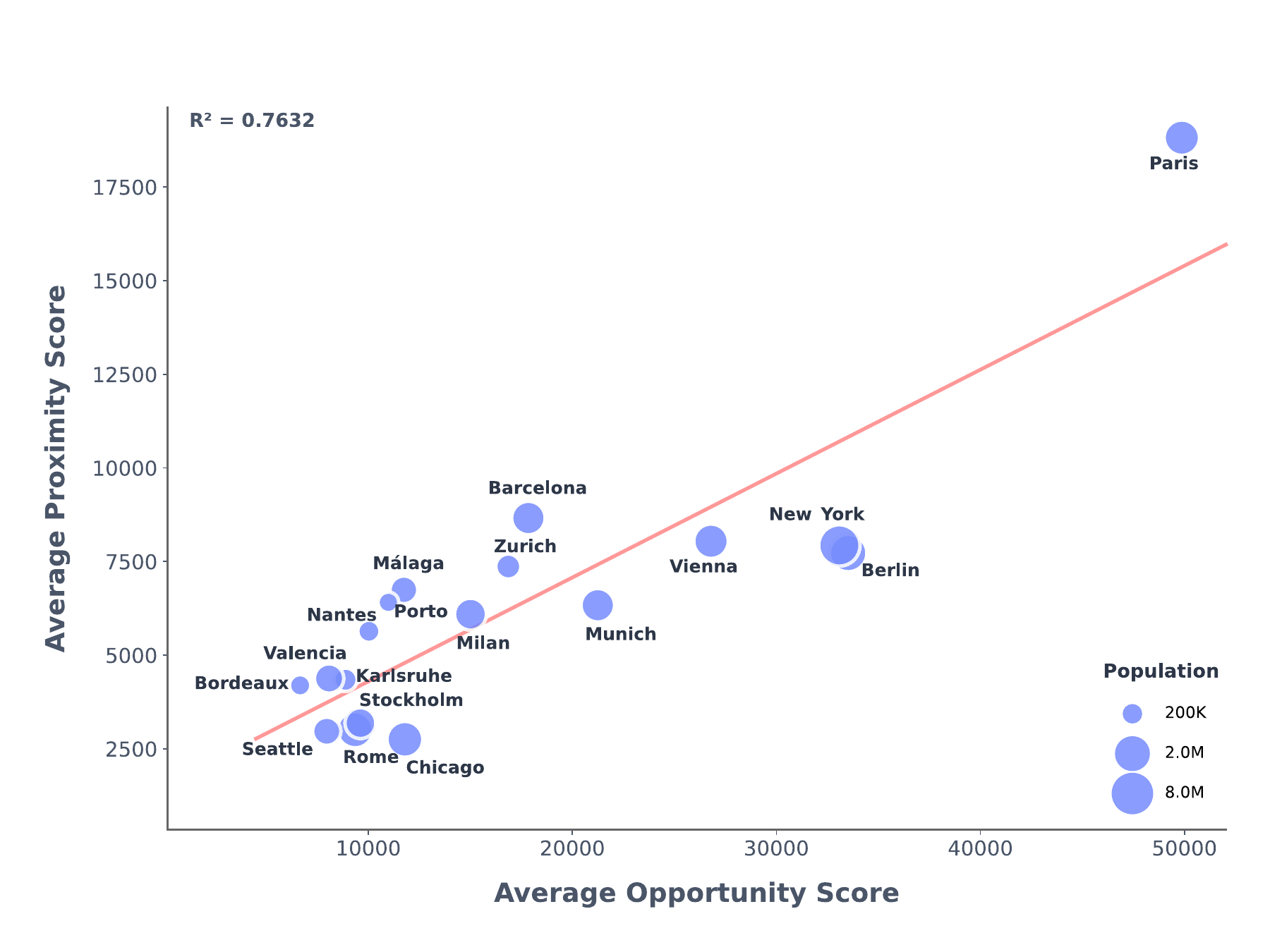}
\caption{\textbf{Diagram with dimensions of accessibility for many cities, averaged.} Cities offer different average experiences, with some offering a typical more isolated life and some having an overall good access. The linear relationship (R² = 0.76) reveals that proximity and opportunity dimensions are strongly correlated at the city level. The average experience, however, is usually the result of heterogeneous access experiences.}\label{fig3:many_cities}
\end{figure}

Taking a population-weighted average for each city, we can calculate a single score for the two accessibility metrics, as in the scatter plot of Fig.~\ref{fig3:many_cities}. The scores differ for each city, but are still correlated, with denser cities allowing for more proximity and opportunity. Paris stands out as the city with the best average access scores in both dimensions, even though the administrative boundaries of Paris are restricted to a very dense and central area of the larger metropolitan area. Two US cities stand in the bottom left area, while New York stands out among the best with Berlin, Vienna, Munich and Barcelona.
\begin{figure}[h]
\centering
\includegraphics[width=0.7\textwidth]{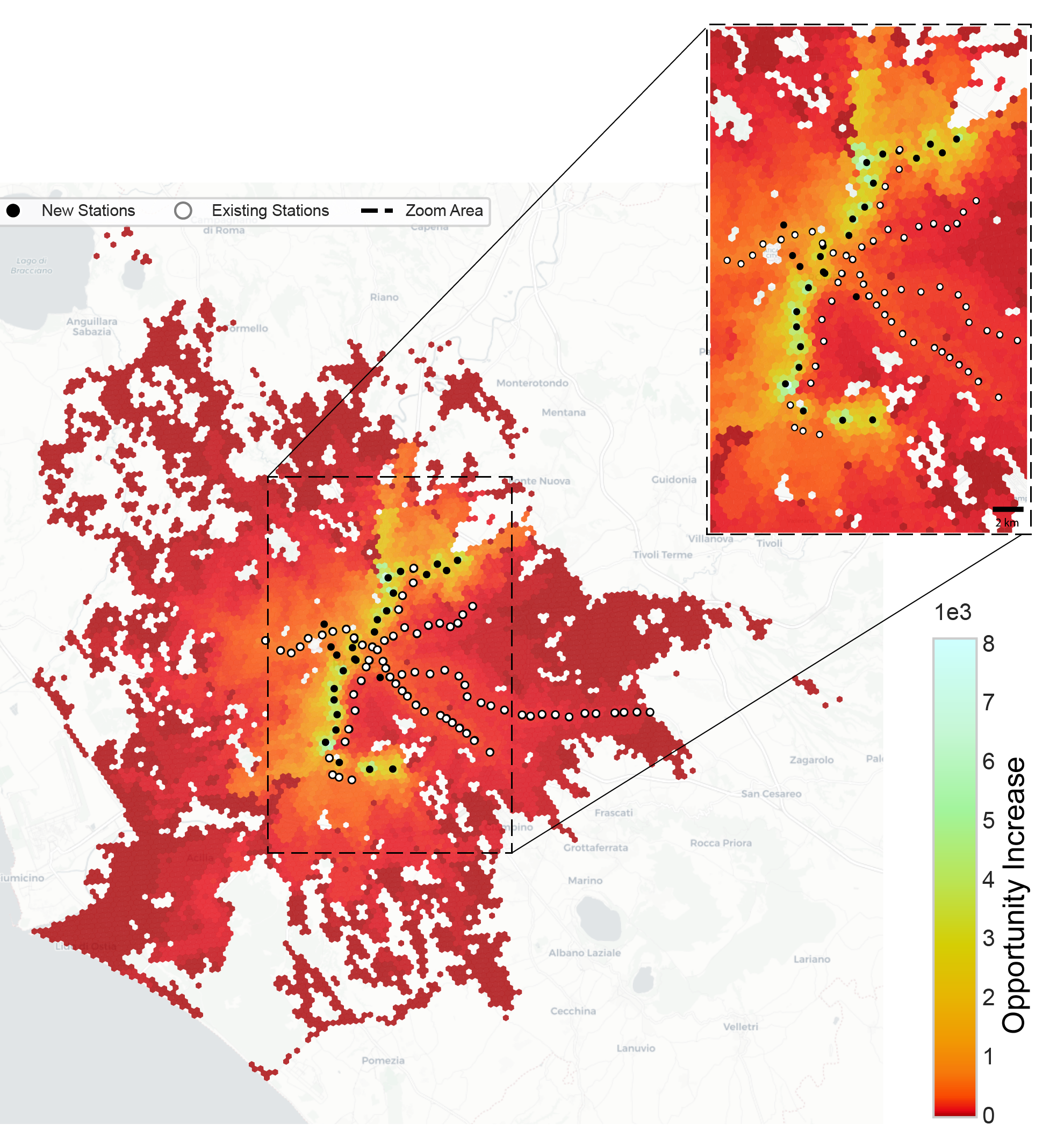}
\caption{\textbf{Scenario implementing planned metro lines in Rome.} Adding the D Line and the last part of the planned C Line to Rome's metro network increases the Opportunity scores for a large part of the city.
}\label{fig4:rome_scenario}
\end{figure}
\rev{A simple least-squares fit across the 18 cities confirms this pattern, yielding a strong positive association between the two city-level scores (with $R^2 \simeq 0.76$), so that cities with higher proximity also tend to enjoy higher opportunity. Although New York and Berlin show comparatively large deviations from the best-fit line, analysing their residuals as a function of basic descriptors of the population density field and related urban indicators did not reveal any robust single factor that could convincingly explain their distance from the regression.}

When planning public transportation or land use changes, we can simulate such changes to see the impact on different areas. In the scenario depicted in Figure \ref{fig4:rome_scenario}, we simulate how two new planned lines would improve opportunity in Rome. This scenario is obtained by implementing the rest of a planned metro line (Linea C\footnote{https://metrocspa.it/lopera/sviluppi-futuri/}) and an additional one that has been inserted in the city's master plan (Linea D)\footnote{https://www.romametropolitane.it/articolo.asp?CodMenu=4\&CodArt=33}. By improving public transportation, the population-weighted opportunity increases from 9353 to 10120, an 8\% increase. 20\% of the residents experience an increase of at least 1000 opportunities, with the areas of Magliana Nuova and Salario increasing their opportunity access by more than 8000.

\section{Discussion and conclusions}\label{sec:discussion}

\rev{In this work, we built on previous research frameworks on accessibility to formalise an accessibility quantification that has 3 axes. With respect to previous research, we populate the node-place diagram of accessibility employing independent spatial accessibility metrics based on access times to POIs by walking (Proximity, for places) and by public transport (Opportunity, for nodes), representing local abundance of opportunities and the possibility to reach far ones in short times. We also add a new axis called Value, that remains an elusive property of urban spaces and therefore is not quantified here but only theorised.}


By singling out \emph{Proximity}, \emph{Opportunity}, and \emph{Value}, \rev{we reformalise and extend earlier two-dimensional frameworks of accessibility, providing an additional lens through which to approach the long-standing debate on how to measure accessibility and evaluate the potential of places}. Classic land‑use models already recognised that distance alone is an imperfect proxy for spatial welfare~\cite{hansen1959accessibility, handy1997measuring,geurs2004accessibility}. More recent conceptual work on proximity‑centred accessibility echoes this view, emphasising that “being near” is necessary but not sufficient if the surrounding opportunities are sparse or misaligned with residents’ needs~\cite{geurs2004accessibility, hill2024beyond, brunoUniversalFrameworkInclusive}. Our tri-axial formulation builds on earlier two-dimensional models by formalising the idea that accessibility depends not only on the ease of reaching desirable destinations, whether near or far, but also on the perceived utility opportunities hold for individuals or communities.

The accessibility literature thus sits on a knife‑edge between communicability and completeness. Geurs and van Wee’s work~\cite{geurs2004accessibility} already warned that every step toward a simpler indicator trades explanatory depth for ease of uptake. This point is reinforced by the works of El-Geneidy and Levinson~\cite{el2006access,el2011place}. After experimenting with the recursively weighted “Place Rank” index, they found that planners and the public gravitated to the headline clarity of cumulative-opportunity counts 
and that this clarity accelerated real‑world adoption\footnote{https://www.transportist.net/p/the-accessibility-turn}.
The paradox is evident: the more intuitive the metric, the less dimensionality it can carry, yet multidimensionality is precisely what makes accessibility a robust diagnostic of urban equity. A tri‑dimensional approach proposes a reduction to this tension by layering \emph{Proximity}, \emph{Opportunity}, and \emph{Value} rather than collapsing them into a single scalar. Coincidentally, the initials of this shift of perspective make a new P.O.V. (Point Of View) in quantifying access. This description preserves simplicity at the communication layer (each axis can still be represented in concrete units). At the same time, theoretical richness is retained in the analytical layer, where the three axes interact to reveal trade‑offs invisible to any one‑dimensional score.

\MOD{From a time-geographical perspective in the sense of Hägerstrand~\cite{hagerstrand1970people}, our three dimensions describe potential access once basic space–time constraints are fixed, while institutional and temporal restrictions, such as opening hours, eligibility rules or monetary fees, can be represented as additional filters on the opportunity set or as shifts in the value dimension, in line with the unified treatment of accessibility proposed by Levinson and co-authors~\cite{levinson2020towards}.}

Operationalising our tri‑dimensional framework begins with pinning down what each axis actually measures. In this regard, our work encounters some limitations that need to be assessed in further research. Proximity is the very short‑range ease of reaching everyday essentials, grocery shops, primary schools, pharmacies, typically captured with a step or decay function such as “within 15 minutes on foot, bike, or transit.” Even this seemingly simple metric can wobble: many Global‑South cities lack complete pedestrian‑network data, reliable speed profiles, or up‑to‑date POI inventories~\cite{liu2022generalized,fry2020assessing,haklay2010good,barron2014comprehensive}.

Opportunity refers to tapping into city‑ or region‑scale assets, diverse job markets, universities, stadiums, and longer but acceptable travel windows (e.g., 15–60 minutes by transit). Measuring it demands high‑resolution, multimodal travel‑time matrices and harmonised land‑use or labour‑market datasets. Yet, GTFS feeds are patchy, informal transit is under‑mapped, and job counts are often aggregated at coarse administrative units, making opportunity scores for cities in the global south potentially far less precise~\cite{williams2015digital,hu2019measuring}.

Only after recognising these baseline data frictions can one tackle the third axis, Value, i.e., the quality or desirability that residents attach to the destinations they can reach. Here, the toolbox widens: crowd‑sourced ratings (Google, Yelp), transaction prices, social‑media check‑ins, public‑health inspections, and bespoke stated‑preference surveys all serve as proxies~\cite{khan2022exploring, rahimi2018geography, schomberg2016supplementing, howie2010application,benati2024unequal}. Each source, however, carries bias. Digital participation skews toward wealthier, younger users; price signals confound quality with ability to pay; survey coverage is costly and localised~\cite{bright2018geodemographic,rao2005quality}. A pragmatic route is to triangulate: weight crowd ratings by review density, normalise price‑based measures by income, and splice in participatory mapping where digital footprints are thin. The result is not a single “value index” but a layered surface that acknowledges uncertainty while illuminating which neighbourhoods enjoy high‑quality choices and which are stranded in “value deserts.” In short, precise measurement across all three dimensions is possible, but only by blending multiple data streams and openly reporting the confidence intervals that inevitably widen outside well‑instrumented, high‑income contexts.

Moreover, when categorising places and cities based on accessibility metrics, we inevitably make several behavioural assumptions. Where do we draw the line between isolation and accessibility? How many opportunities are sufficient or desirable in one’s surroundings and life, and what is the tradeoff with the resulting density supporting this abundance? And besides, are more opportunities always better? If not, how does a “paradox of choice”~\cite{schwartz2015paradox} apply to the fast-paced reality of urban life? Personal preferences likely shape these questions, and, fortunately, cities remain diverse environments where different areas offer different experiences. However, the line between heterogeneity and inequality is difficult to define and measure. For this reason, when quantifying access in cities, we should avoid assuming that “more is always better”. Aggregate scores of local metrics can be misleading and conceptually flawed when taken at face value.


Finally, taken together, the evidence and arguments above position accessibility as a living, three‑dimensional field, whose contours evolve through continual feedback between land use, transport, and human preference. While simple, single-number indicators may be tempting, they often obscure the crucial trade-offs that affect equity and resilience. By implementing each aspect with clear, multi-source data and incorporating interventions into models that are sensitive to feedback, we can shift from basic assessments to truly adaptive planning. The task ahead is not to discover one perfect metric, but to cultivate a toolkit that lets cities track, simulate, and steer accessibility as it changes, ensuring that improvements in one dimension never come at the unseen expense of another.

\backmatter


\bmhead{Ethics approval and consent to participate}
\MOD{We confirm that no ethical concerns are applicable to this paper and that all the authord agreed to participate.}

\bmhead{Consent for publication}
\MOD{We confirm the corresponding author has read the journal policies and submit this manuscript in accordance with those policies.}

\bmhead{Availability of data and materials}
\MOD{All data used in this study are publicly available: administrative boundaries from municipal open data portals, public transport schedules in GTFS format from transport agency websites (downloaded May 2025), Points of Interest from OpenStreetMap (https://www.openstreetmap.org), and population density grids from WorldPop (https://www.worldpop.org).}

\bmhead{Competing interests}
\MOD{The authors declare that ave no competing interests as defined by Springer, or other interests that might be perceived to influence the results and/or discussion reported in this paper.}

\bmhead{Funding}
\MOD{This research did not receive funding.}

\bmhead{Authors' contributions}
\MOD{Research design and study concept: All authors. Data analysis: M.B., B.C., L.R.M.. Result interpretation: all authors. Manuscript drafting, review and editing: all authors.}

\bmhead{Acknowledgements}
\MOD{H.P.M.M. acknowledges the support of Fundação Edson Queiroz, Universidade de Fortaleza, and Fundação Cearense de Apoio ao Desenvolvimento Científico e Tecnológico. Basemaps for Fig. \ref{fig2:ny_case_study} and \ref{fig4:rome_scenario} © CARTO, © OpenStreetMap contributors, https://www.openstreetmap.org/copyright.}

\bibliography{bibliography}


\end{document}